\documentclass{pasj00} 
\Received{$\langle$reception date$\rangle$}
\Accepted{$\langle$acception date$\rangle$}
\Published{$\langle$publication date$\rangle$}
\SetRunningHead{Prospects for Very High Energy Blazar Survey by the
  Next Generation Cherenkov Telescopes}{Y. Inoue, T. Totani, \&
  M. Mori}

\usepackage{times}
\usepackage{lscape}

\begin{document}
\title{Prospects for Very High Energy Blazar Survey by the Next
  Generation Cherenkov Telescopes} 

\author{Yoshiyuki
  Inoue\altaffilmark{1}, Tomonori Totani\altaffilmark{1}, \& Masaki
  Mori\altaffilmark{2}} 

\affil{$^1$Department of Astronomy, Kyoto
  University, Kitashirakawa, Sakyo-ku, Kyoto 606-8502, Japan}
\affil{$^2$Department of Physics, Ritsumeikan University, 1-1-1 Noji
  Higashi, Kusatsu, Shiga 525-8577, Japan}

\email{yinoue@kusastro.kyoto-u.ac.jp}


\maketitle
\begin{abstract}
  The prospects for future blazar surveys by next-generation very high
  energy (VHE) gamma-ray telescopes, such as Advanced Gamma-ray
  Imaging System (AGIS) and Cherenkov Telescope Array (CTA), are
  investigated using the latest model of blazar luminosity function
  and its evolution which is in good agreement with the flux and
  redshift distribution of observed blazars as well as the
  extragalactic gamma-ray background.  We extend and improve the
  template of spectral energy distributions (SEDs) based on the blazar
  SED sequence paradigm, to make it reliable also in the VHE bands (above 100 GeV) by
  comparing with the existing VHE blazar data. Assuming the planned
  CTA sensitivities, a blind survey using a total survey time of $\sim
  100$ hrs could detect $\sim 3$ VHE blazars, with larger expected
  numbers for wider/shallower surveys. We also discuss following-up of
  {\it Fermi} blazars. Detectability of VHE blazars in the plane of {\it Fermi}
  flux and redshift is presented, which would be useful for future
  survey planning. Prospects and strategies are discussed to
  constrain the extragalactic background light (EBL) by using the
  absorption feature of brightest blazar spectra, as well as cut-offs
  in the redshift distribution. We will be able to get useful
  constraints on EBL by VHE blazars at different redshifts ranging
  0.3--1 TeV corresponding to $z=$0.10--0.36.
\end{abstract}

\section{Introduction}
\label{sec:intro}

Very high energy (VHE; above 100 GeV) gamma-ray astronomy has now
firmly been established by the observations of the state-of-the-art
imaging atmospheric Cherenkov Telescopes (IACTs) such as H.E.S.S.,
MAGIC, and VERITAS (see {de Angelis}, {Mansutti}, \&  {Persic} 2008; Mori 2009, for
reviews).  Further progress is
anticipated in the near future by the planned next-generation IACTs
such as Cherenkov Telescope Array (CTA) and Advanced Gamma-ray Imaging
System (AGIS).  The sensitivities of all-sky monitoring VHE gamma-ray
experiments are also expected to improve by the future projects such
as the High Altitude Water Cherenkov Experiment (HAWC) and the
Tibet-III/MD experiment.

Current IACTs have already found $\sim$ 100 VHE sources including
$\sim$ 25 blazars. Blazars, a class of active galactic nuclei (AGNs),
are the dominant population in the extragalactic gamma-ray sky. Almost
all of the extragalactic sources detected by EGRET (Energetic
Gamma-Ray Experiment Telescope) on board the Compton Gamma Ray
Observatory are blazars ({Hartman} {et~al.} 1999). Moreover, 3
months bright source and 11 months catalog by the Fermi gamma-ray
space telescope ({\it Fermi}) have recently also showed that most of
the extragalactic sources are blazars
({Abdo} {et~al.} 2009a, 2009c, 2010), and we expect
that more than 1000 blazars will be detected by {\it Fermi} in the
near future
(e.g. {Narumoto} \& {Totani} 2006; {Dermer} 2007; {Inoue} \& {Totani} 2009).
The number of VHE blazars are expected to dramatically increase with
the improved next-generation IACT sensitivity. Therefore, it would be
possible to do a statistical study of VHE blazars in the CTA/AGIS era,
which would provide a crucial key to understand AGN populations and
high energy phenomena around super massive black holes in AGNs and
jets.

The purpose of this paper is to study the prospect of future blazar
surveys by IACTs, especially for the statistical power of future VHE
blazar sample that can be obtained by realistic observing time of
next-generation IACTs.  For this purpose, blazar gamma-ray luminosity
function (GLF) and spectral energy distribution (SED) are needed. The
blazar GLF has been studied in detail by many papers
({Padovani} {et~al.} 1993; {Stecker}, {Salamon}, \&  {Malkan} 1993; {Salamon} \& {Stecker} 1994; {Chiang} {et~al.} 1995; {Stecker} \& {Salamon} 1996; {Chiang} \& {Mukherjee} 1998; {M{\"u}cke} \& {Pohl} 2000; {Narumoto} \& {Totani} 2006; {Dermer} 2007; {Inoue} \& {Totani} 2009). {Inoue} \& {Totani} (2009)
(hereafter IT09) has recently presented a new blazar GLF taking into
account the blazar SED sequence (see \S \ref{subsec:glf}), which is in
nice agreement with the CGRO/EGRET and Fermi/LAT data.
We utilize this IT09 model to predict the expected number and
distributions of physical quantities of VHE blazars in future IACT
surveys. Since the SED model of IT09 was constrained only at the
photon energies under GeV, we construct a new blazar SED template by
modifying that used in IT09 in accordance with the available VHE
blazar data. By using our updated blazar sequence and GLF model, it is
possible for us to make predictions for future VHE gamma-ray
observations, which is the most reliable based on available observed
data.

Extragalactic background light (EBL) in the optical and infrared bands
contains the information about the history of star formation activity
in the universe, and knowing EBL quantitatively is an important step
to understand galaxy formation in the cosmological context.  However,
it is hard to measure EBL spectrum directly, mainly because of the
difficulty in subtracting foreground emission (see {Hauser} \& {Dwek} 2001, for
reviews). VHE observations provide a completely
independent constraint on EBL, since VHE gamma-ray photons propagating
the universe are absorbed via electron-positron pair creation with the
EBL photons ({Gould} \& {Schr{\'e}der} 1966; {Jelley} 1966). Some
useful limits have already been obtained by VHE blazar observations
({Aharonian} {et~al.} 2006a; {MAGIC Collaboration}, {Albert},  {et~al.} 2008), up to the redshift of
$z=0.536$ by using 3C279 data.  The next generation IACTs will shed further
light on this issue, and we discuss the prospect about this as a
particular application of our study.

This paper is organized as follows. We introduce our updated blazar
SED template and GLF model, as well as the model of VHE gamma-ray
absorptions by EBL in \S \ref{sec:model}. In \S \ref{sec:cta}, we make
predictions for the expected number and statistics of future VHE
blazar surveys assuming some observing modes.  We discuss the
prospect for the determination of EBL by VHE blazars in \S
\ref{sec:dis}. Summary is given in \S \ref{sec:sum}. Throughout this
paper, we adopt the standard cosmological parameters of
($h,\Omega_M,\Omega_\Lambda$)=(0.7,0.3,0.7).

\section{Model Description}
\label{sec:model}

\subsection{Blazar Gamma-ray Spectrum and Luminosity Function}
\label{subsec:glf}

\begin{figure*}
\begin{center}
\FigureFile(150mm,150mm){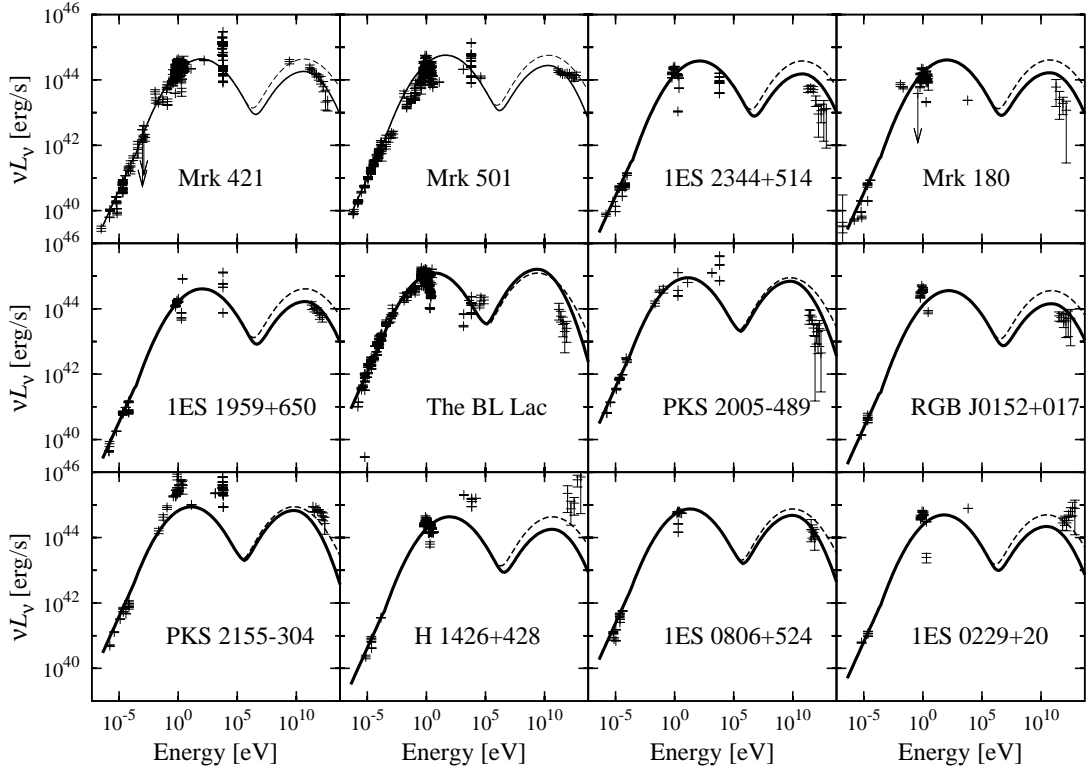} 
\end{center}
\caption{ SEDs (in isotropic equivalent luminosity) of VHE blazars at
  $z \leq 0.14$.  The VHE data points are taken from the literature:
  Mrk 421({Albert} {et~al.} 2007c), Mrk
  501({Albert} {et~al.} 2007d), 1ES
  2344+514({Albert} {et~al.} 2007b), Mrk
  180({Albert} {et~al.} 2006a), 1ES 1959+650
  ({Albert} {et~al.} 2006b), the BL Lac ({Albert} {et~al.} 2007a),
  PKS 2005-489
  ({Aharonian} {et~al.} 2005a), RGB J0152+017
  ({Aharonian} {et~al.} 2008), PKS 2155-304
  ({Aharonian} {et~al.} 2005b), 1ES 0806+524
  ({Acciari} {et~al.} 2009), H 1426+428 ({Aharonian} {et~al.} 2002),
  and 1ES 0229+200 ({Aharonian} {et~al.} 2007). Note that VHE data are
  deabsorbed by the EBL model of {Totani} \& {Takeuchi} (2002).  The
  data points at energy bands other than VHE are taken from NED. Solid
  and dashed curves correspond to blazar sequence models by this paper
  and IT09, respectively.
  \label{fig:sed_multi} }
\end{figure*}

IT09 has recently developed a blazar GLF model based on the latest
determination of X-ray luminosity function of AGNs
({Ueda} {et~al.} 2003; {Hasinger}, {Miyaji}, \&  {Schmidt} 2005), featuring so called
luminosity dependent density evolution (LDDE). Another new feature of
IT09 is taking into account the blazar SED sequence. Blazar sequence
is a feature seen in the mean SED of blazars that the synchrotron and
inverse Compton (IC) peak photon energies decrease as the bolometric
luminosity increases
[{Fossati} {et~al.} (1997); {Kubo} {et~al.} (1998); {Fossati} {et~al.} (1998); {Donato} {et~al.} (2001); {Ghisellini}, {Maraschi}, \&  {Tavecchio} (2009),
but see also {Padovani} {et~al.} (2007)].  The key parameters in GLF
have carefully been determined to match the observed flux and redshift
distribution of EGRET blazars by a likelihood analysis.  Recently, the
predicted extragalactic gamma-ray background (EGRB) spectrum by IT09
including non-blazar AGNs contributing to MeV bands
({Inoue}, {Totani}, \&  {Ueda} 2008) has been found to be in excellent
agreement with the new determination of the EGRB spectrum reported by
{\it Fermi} (Fermi-LAT collaboration 2010; {Inoue} {et~al.} 2010).

\begin{figure}
\begin{center}
\FigureFile(90mm,90mm){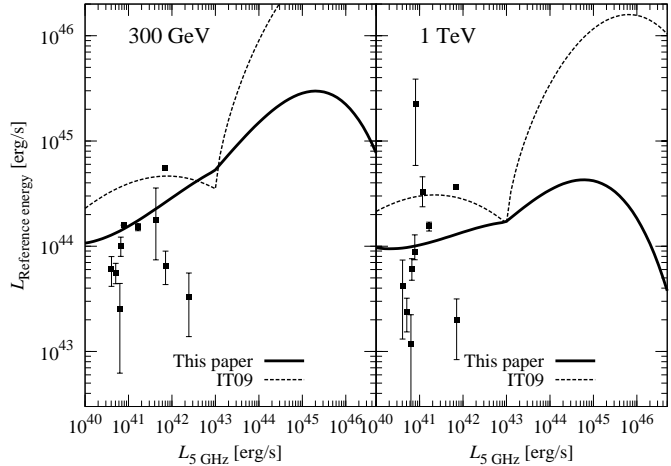} 
\end{center}
\caption{
Gamma-ray and 5GHz radio luminosity relations in $\nu L_\nu$.
The VHE luminosities in the left and right panels are at 300 GeV and
1 TeV, respectively.  Data points are the same as those of 12
  blazars in Fig. \ref{fig:sed_multi}, except for H 1426+428 
and 1ES 0229+200 in the left panel, and 
the BL Lac and 1ES 0806+524 in the right panel,
because of non-detections in the corresponding energy bands.
  Gamma-ray luminosity is deabsorbed using the
  optical depth model of {Totani} \& {Takeuchi} (2002) for
  intergalactic absorption.  Solid and dashed curves correspond to
  the blazar sequence models by this paper and IT09,
  respectively. 
  \label{fig:Lr_LTeV} }
\end{figure}

The gamma-ray SED of the blazar sequence model used in IT09 is
constrained only by the EGRET data whose energy range is 30 MeV -- 30
GeV. Figs. \ref{fig:sed_multi} and \ref{fig:Lr_LTeV} show
multi-wavelength SEDs and radio-to-gamma-ray luminosity relation of
VHE blazars, respectively. To avoid the absorption effect by EBL at
high redshift, we have selected 12 VHE blazars below $z=0.14$ where
optical depth for 1 TeV photon is $\lesssim1$. We have obtained SED data
from published papers for VHE gamma-ray data (see the caption of
Fig. \ref{fig:sed_multi}) 
and from the NASA/IPAC
Extragalactic Database (NED) for other wavelength data\footnote{NED is
  operated by the Jet Propulsion Laboratory, California Institute of
  Technology, under contract with the National Aeronautics and Space
  Administration.}.  By comparing with the observed data of VHE
blazars, we have updated our blazar SED sequence model to properly
reproduce typical VHE flux of observed blazars.  A source of
systematic uncertainty in this procedure is the variability of
blazars; generally blazars show rapid and violent variability, and
hence it is difficult to accurately estimate the VHE luminosity
averaged over a long time. Here we simply collected published VHE flux
data in the literature, except for those of observations aiming at
blazars during flares.

As shown in Figs. \ref{fig:sed_multi} and \ref{fig:Lr_LTeV}, the
sequence model of IT09 tends to overestimate VHE luminosity, but the
new formulation reproduces a rough mean of VHE luminosities, though
there is still significant scatter around the mean. Furthermore, the
IT09 model shows a kink in the radio-VHE gamma-ray luminosity
correlation, because of a mathematical problem of the connection
between different luminosity ranges.  This kink has also been removed
in the new formulation.  The new formulation of our updated blazar SED
sequence templates is presented in Appendix in detail. We use this
sequence template as the best model currently available to predict the
statistics of VHE blazars based on the luminosity function determined
at lower photon energy bands. Our SED template is also consistent with very recent MAGIC stacking
analyzed BL Lac SED (MAGIC Collaboration 2010)

We have also reconstructed the blazar GLF model based on our modified
blazar sequence formulation.  We set minimum and maximum gamma-ray
luminosities of blazars as $10^{43} \rm erg/s$ and $10^{50} \rm erg/s$
in $\nu L_\nu$ at rest-frame 100 MeV as in IT09 \footnote{Throughout
  this paper, the blazar luminosity is expressed by
  isotropic-equivalent luminosity, though blazar emission should be
  strongly beamed in reality.}.  Since the modification in SED is
mostly in the VHE energy band, the predictions for other wavelength,
including the GeV energy band for {\it Fermi}, hardly change from
IT09.  For example, the expected {\it Fermi} blazar count is $\sim
720$ and $\sim 750$ in the entire sky for our model and IT09,
respectively, where we set the {\it Fermi} sensitivity as $3 \times
10^{-9} \ \rm photons \ cm^{-2} s^{-1}$ at $>$100 MeV corresponding to
the 1-year sky survey sensitivity ({Atwood} {et~al.} 2009).  Since {\it
  Fermi} 11-months AGN catalog has already detected 596 blazars and 72
unidentified extragalactic gamma-ray sources at high Galactic latitude
($|b|>10^\circ$), our source count prediction is also consistent with
the number of {\it Fermi} blazars. 

The key parameters of the blazar GLF are $(q,\ \gamma_1, \kappa)=(4.50, 1.10,
1.42\times10^{-6})$, where $q$ is the ratio between the bolometric jet
luminosity and disk X-ray luminosity, $\gamma_1$ the faint-end slope
index of GLF, and $\kappa$ a normalization factor of GLF (see
Section. 3 of IT09 for details).  Now we can predict the abundance and
statistics of blazars in any photon energy bands including VHE
gamma-ray.

\subsection{EBL models}
\label{subsec:ebl}

\begin{figure*}
\begin{center}
\FigureFile(150mm,150mm){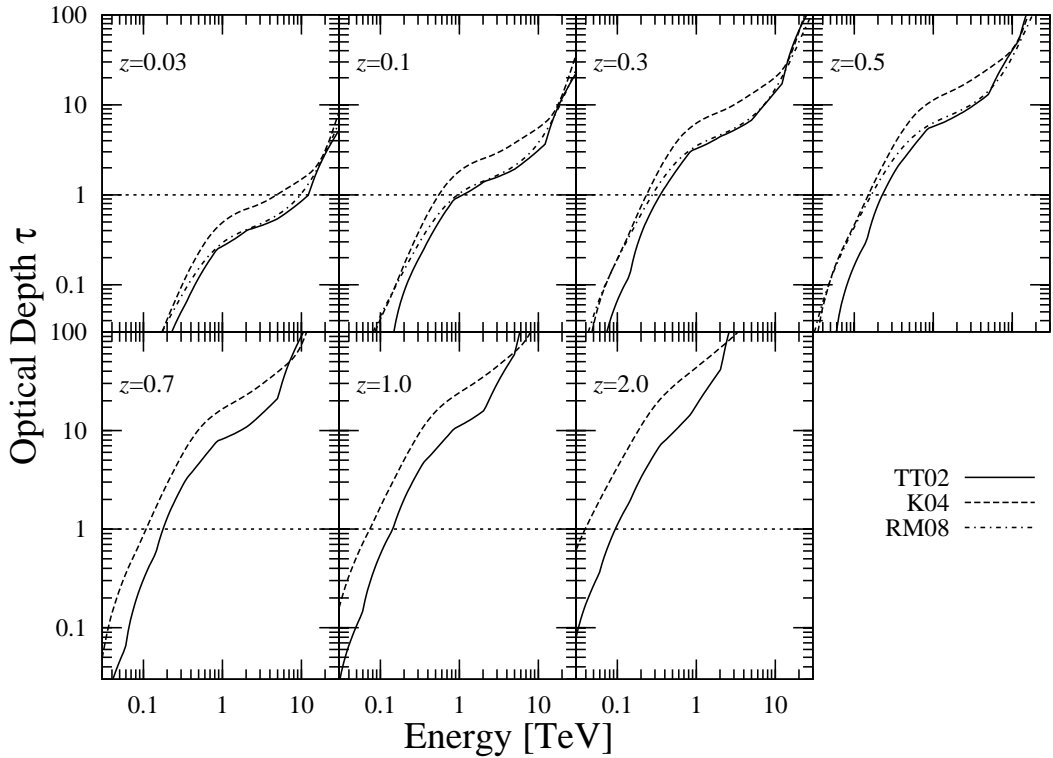} 
\end{center}
\caption{Optical depth of intergalactic absorption
of high energy gamma-rays for various source redshifts as indicated in each
  panel. Solid, dashed, and dot-dashed curves corresponds to 
the models of
{Totani} \& {Takeuchi} (2002, TT02), {Kneiske} {et~al.} (2004, K04), and {Raue} \& {Mazin}(2008, RM08), respectively. The dotted line marks the level of
  optical depth $\tau=1$.\label{fig:tau} 
}
\end{figure*}

When we try to make some
predictions for extragalactic VHE blazar survey, EBL modeling is
crucial.  A number of models have been proposed by many authors
(e.g. {Salamon} \& {Stecker} 1998; {Totani} \& {Takeuchi} 2002; {Kneiske}, {Mannheim}, \&  {Hartmann} 2002; {Kneiske} {et~al.} 2004; {Primack}, {Bullock}, \&  {Somerville} 2005; {Stecker}, {Malkan}, \&  {Scully} 2006; {Mazin} \& {Raue} 2007;Raue \& Mazin(2008)). Fig. \ref{fig:tau} shows optical depth
models of {Totani} \& {Takeuchi} (2002, TT02), {Kneiske} {et~al.} (2004, K04), and {Raue} \& {Mazin}(2008, RM08). Since RM08
constrained EBL optical depth from VHE blazars observations, we
present their model below $z=0.6$. 
Here we use the 
optical depth of TT02 as the standard in this paper,
because it is in good agreement with RM08 that is consistent
with VHE blazar observations at low redshift, and it extends beyond $z \sim 1$
by galaxy evolution modeling that is consistent with galaxy counts and
EBL observations in optical and infrared bands (TT02). 
K04 predicted about twice higher optical depth
than TT02, and the expected number of VHE blazars will decrease from
our predictions below, when we adopt K04.

\section{Predictions for the upcoming CTA era}
\label{sec:cta}

We consider two modes of future surveys for VHE blazars.  One is a
blank field sky survey.  This is a natural outcome for the nearly
all-sky monitoring types of VHE observatories (such as the HAWC and
Tibet experiments).  For IACTs, various survey designs are possible
for a fixed amount of the total observation time, changing the survey
area and exposure time for one field of view (FoV) [e.g., the Galactic
Plane by H.E.S.S. ({Aharonian} {et~al.} 2006b)].  The other is
follow-up surveys for targets selected at other wavelengths. We
particularly consider a follow-up survey of Fermi blazars by CTA.

The sensitivity of VHE detectors is often given in terms of
integrated photon flux (total photon flux above a given photon
energy).  However, a comparison with theoretical prediction is more
easily made in terms of energy flux like $\nu F_\nu$.  In this paper,
we express the VHE sensitivity in terms of $\nu F_\nu=E^2
dF_\gamma/dE$, where $E$ is the gamma-ray energy and $dF_\gamma/dE$
the differential photon flux.  The sensitivities in integrated photon
flux of H.E.S.S., CTA , Tibet-III/MD, and HAWC\footnote{H.E.S.S.:
  {http://www.mpi-hd.mpg.de/hfm/HESS/}\\CTA:
  {http://www.cta-observatory.org/}\\ Tibet-III/MD:
  {http://www.icrr.u-tokyo.ac.jp/em/index.html}\\ HAWC:
  {http://hawc.umd.edu/}} are converted into $\nu F_\nu$ assuming a
gamma-ray spectrum of $dF_\gamma/dE\propto E^{-2.5}$, although
spectral index varies from source to source in reality. We also assume
that the sensitivity limit scales as $\propto T^{-1/2}$, where $T$ is
the exposure time.  Table \ref{tab:sens} summarizes the 5$\sigma$ CTA
sensitivities in several energy bands for some sets of observing time.
The sensitivity of CTA will be about one order of magnitude better
than that of current IACTs, and the photon energy range will also
become about one order of magnitude wider.

\begin{table}
 \begin{center}
  \caption{CTA $\nu F_\nu$ sensitivity in the unit of $10^{-13}\ \rm erg/cm^2/s$\footnotemark[$*$]. }\label{tab:sens}
   \begin{tabular}{lccc}
      \hline
      &   \multicolumn{3}{c}{Observing time per 1 FoV} \\ \cline{2-4}
      Energy & 2 hrs & 10 hrs & 50 hrs \\ \hline
		30 GeV & 45 & 20 & 9.0 \\
		100 GeV & 25 & 11 & 5.0 \\
		300 GeV & 5.0 & 2.2 & 1.0\\
		1 TeV & 3.0 & 1.3 & 0.6\\
		10 TeV & 10 & 4.5 & 2.0\\ \hline
       \multicolumn{4}{@{}l@{}}{\hbox to 0pt{\parbox{70mm}{\footnotesize
       \par\noindent
       \footnotemark[$*$] 
These sensitivities are converted into $\nu F_\nu$ basis
from those in integrated photon flux of 5-$\sigma$, 50 hrs
observation ({http://www.cta-observatory.org/}). 
Sources are assumed to
have a power law differential photon spectrum 
of $dF_\gamma/dE \propto E^{-2.5}$.
}\hss}}
    \end{tabular}
  \end{center}
\end{table}

\subsection{Blank Field Surveys}
\label{subsec:bfs}

\begin{figure*}[t]
\begin{center}
\FigureFile(140mm,140mm){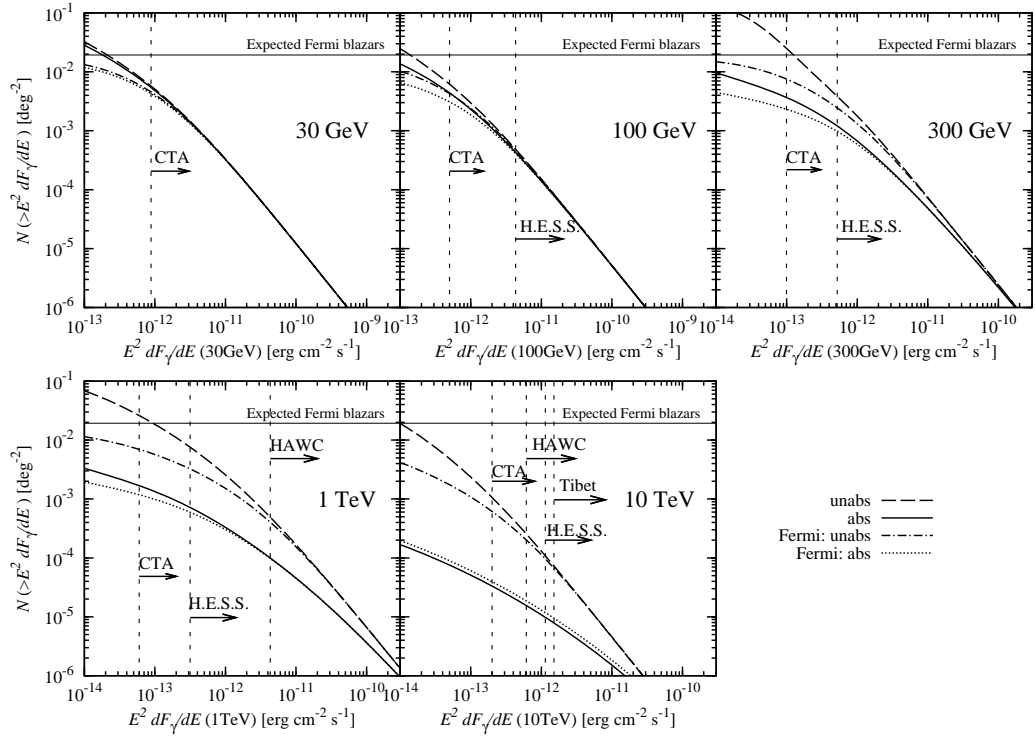} 
\end{center}
\caption{
Cumulative source counts as a function of gamma-ray flux (in $\nu
F_\nu$) of VHE blazars.  The five panels correspond to different
photon energies as indicated in the panels.
The solid curves are predictions by our
blazar GLF model.  The dotted curves are the same as the solid curves,
but for blazars that are detected by Fermi with a sensitivity of
$F_{\rm lim} =3\times 10^{-9}$ photons cm$^{-2}$ s$^{-1}$ for photon
flux above 100 MeV.  The intergalactic absorption by EBL is taken into
account for the solid and dotted curve, but not in the dashed and
dot-dashed curve.  
The 5-$\sigma$ detection limits of H.E.S.S. and CTA
for 50-hr observation and those of HWAC and Tibet-III/MD for 1-year
observation are also shown. 
The dotted curve in the panel of 10 TeV
is shifted upward artificially by a factor of 1.2 for the purpose of presentation,
because the solid and dotted
curves totally overlap with each other.  
The horizontal thin
solid line is the total expected number of Fermi blazars
with the Fermi sensitivity given above.
\label{fig:count} }
\end{figure*}

A blank field survey is the most fundamental mode of observing a sky
in a waveband, and free from biases about the pre-selection, except
the flux limit of the survey.  A catalog of objects obtained by such a
survey is important for a statistical study, such as construction of
luminosity function. For example, the Galactic plane survey by
H.E.S.S. made a breakthrough in the Galactic high energy astronomy by
discovering various gamma-ray emitting objects ({Aharonian} {et~al.} 2006b).

Fig. \ref{fig:count} shows the cumulative source counts per 1 square
degree, i.e., the surface number density of blazars brighter than
a given threshold flux,
predicted by our blazar GLF model in five energy bands of 30
GeV, 100 GeV, 300 GeV, 1 TeV, and 10 TeV as indicated in the
panels. The expected counts in the case of no intergalactic absorption
are also shown.  

First we examine the expected number of blazars detectable by
HAWC or Tibet-III/MD experiments. These experiments cover photon
energies higher than $\sim$ 1 TeV, and the 1-yr, 5-$\sigma$
sensitivities of these two telescopes are indicated in the 1 and 10
TeV panels of Fig. \ref{fig:count}.  The expected number of blazars by
a HAWC search at 1 TeV is about four in $4\pi$ steradian, and there
may be a chance to detect some bright blazars by HAWC. On the other
hand, the expected number is less than one at 10 TeV.

Next we consider a blind survey in a fixed survey area $A_{\rm
  survey}$ by multiple CTA pointing observations.  The 50-hr,
5-$\sigma$ sensitivities of CTA are shown in Fig \ref{fig:count}.  As
an example, we consider a total survey time of $T_{\rm survey} = $100
hours, and hence the observation time per field-of-view becomes
$T_{\rm FoV} = T_{\rm survey}(A_{\rm FoV}/A_{\rm survey})$.  Here we
assume the CTA FoV to be $A_{\rm FoV} = $20 deg$^2$
({Aharonian} {et~al.} 2006b). Assuming that the flux sensitivity limit
simply scales as $\propto T_{\rm FoV}^{-1/2}$, we find that the
expected number of blazars are 0.17, 0.56, 0.89, 2.3, and 3.3 for the
assumed survey areas of $A_{\rm survey} = 40, 200, 400, 2000$, and
4000 deg$^2$, 
respectively, in the energy band of 300 GeV.  The
expected numbers in the other energy bands are summarized in Table
\ref{tab:blank}. These results mean that a wider and shallower sky
survey is better for a fixed total survey time. However, the expected
number of detectable blazars in a blind survey is at most a few in the
case of $T_{\rm survey}$ = 100 hrs, which is insufficient for a
detailed statistical study such as luminosity function. Typical total
observable time for IACTs are 1000 hrs in a year
, and a more ambitious survey using $\gtrsim$ 1000 hrs may be required
to construct a sufficiently large sample.  Another important
implication is that the contamination of extragalactic objects will be
small in the Galactic plane survey by CTA.

It should be kept in mind that there is a considerable uncertainties
in the numbers predicted above.  The use of the blazar SED sequence is
the key to convert the blazar luminosity function in the GeV band into
VHE band, but the validity of the blazar sequence is still a matter of
debate.  Furthermore, VHE SED of our new sequence model is constrained
by only 12 VHE blazars. We did not consider the time variability of
blazars, because it is difficult to formulate the variability. The
luminosity function model parameters have been determined only by
about 50 EGRET blazars, but a much larger statistics by {\it Fermi}
will soon allow us a more accurate determination of the parameters of
blazar sequence and GLF. Finally, our model includes only known blazar
population. A completely different extragalactic population may be
found by such a blind survey, which is probably the most exciting
possibility and a strong motivation for the survey.

\begin{table}
 \begin{center}
  \caption{Expected blazar counts for 100 hours CTA blank field survey}\label{tab:blank}
   \begin{tabular}{lccccc}
      \hline
      &   \multicolumn{5}{c}{$A_{\rm survey}$ [deg$^2$]} \\ \cline{2-6}
      Energy  & 40 & 200 &  400 & 2000 & 4000 \\ \hline
		30 GeV & 0.26 & 0.59 & 0.80 & 1.4 & 1.7 \\
		100 GeV & 0.22 & 0.52 & 0.72 & 1.3 & 1.6 \\
		300 GeV & 0.17 & 0.56 & 0.89 & 2.3 & 3.3\\
		1 TeV & 0.05 & 0.14 & 0.23 & 0.61 & 0.92 \\
		10 TeV & 0.002 & 0.004 & 0.007 & 0.02 & 0.03\\ \hline
    \end{tabular}
  \end{center}
\end{table}

\begin{figure*}[t]
\begin{center}
\FigureFile(140mm,140mm){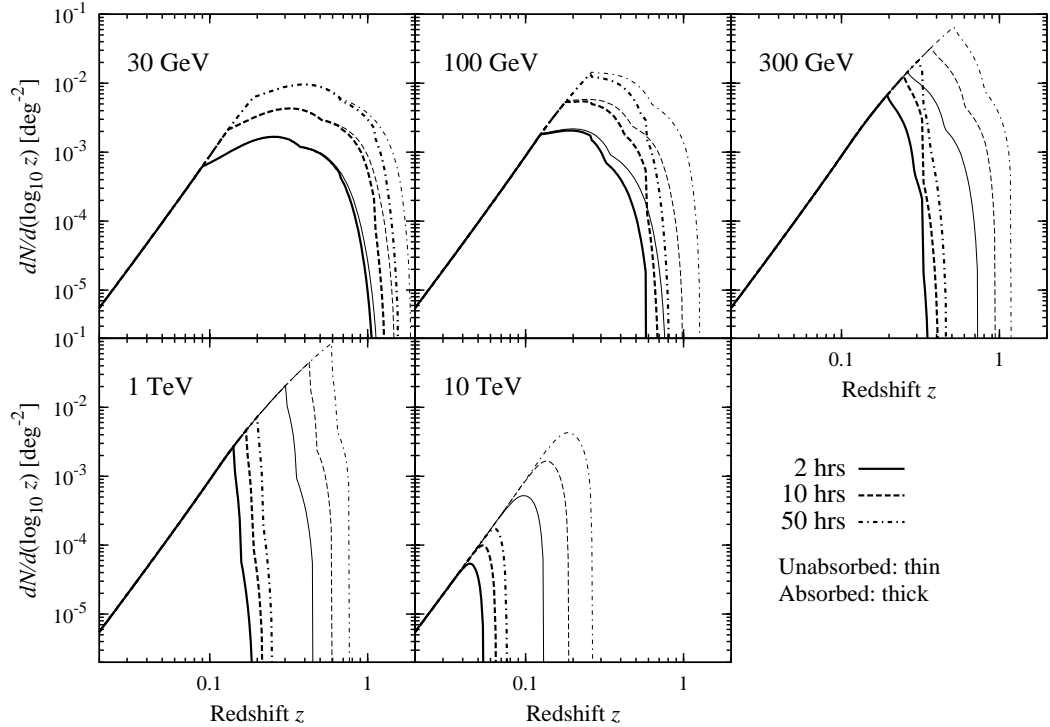}
\end{center} 
\caption{ The differential redshift distribution of VHE blazars in a
  blind survey down to several values of flux sensitivity.  Here, the
  5-$\sigma$ sensitivity is indicated by the corresponding observation
  time per field (see Table \ref{tab:sens} for the sensitivities in
  physical units).  Thick and thin curves are
  predictions when the intergalactic absorption is taken into account
  or not, respectively.
  \label{fig:z_dist} }
\end{figure*}

Fig. \ref{fig:z_dist} shows the expected differential redshift
distribution for several flux limits. Again, the cases of no
intergalactic absorption are also plotted, and the effect of EBL
absorption eliminating high redshift blazars is clearly
seen. Therefore, the number and highest redshift of VHE blazars would
not dramatically increase even with the CTA sensitivity.

\subsection{Following up Fermi blazars}
\label{subsec:fermi}

Since a blind survey for VHE blazars seems not easy or requiring a
large amount of observing time even with the CTA sensitivities, we
consider another strategy to find new VHE blazars, i.e., following up
blazars detected in other wavelengths. The results of the three-month
bright source and 11-month AGN catalog by {\it Fermi} have already
been published ({Abdo} {et~al.} 2009a, 2009c, 2010),
including 596 sources identified as blazars.  It is expected that {\it
  Fermi} will eventually discover $\gtrsim 1000$ blazars in future
survey ({Dermer} 2007; {Inoue} \& {Totani} 2009).  Very recently
{Abdo} {et~al.} (2009b) have reported that {\it Fermi} detected GeV
gamma-rays from 21 blazars that have been detected in TeV bands.
Therefore following up {\it Fermi} blazars would be one of the most
promising ways to increase VHE blazar samples.

Fig. \ref{fig:count} also shows the expected cumulative source counts
of blazars that will be detected by {\it Fermi}, where we set the {\it
  Fermi} sensitivity as $3 \times 10^{-9} \ \rm photons \ cm^{-2}
s^{-1}$ at $>$100 MeV for one-year survey
({Atwood} {et~al.} 2009). We adopt this value for {\it Fermi}
sensitivity throughout the paper unless otherwise stated.  Taking the
sensitivity of 50 hrs observation by CTA (indicated in the figure),
the expected number of detectable blazars is not much different
(within a factor of about two) from that of a blind search.

\begin{figure*}[t]
\begin{center}
\FigureFile(140mm,140mm){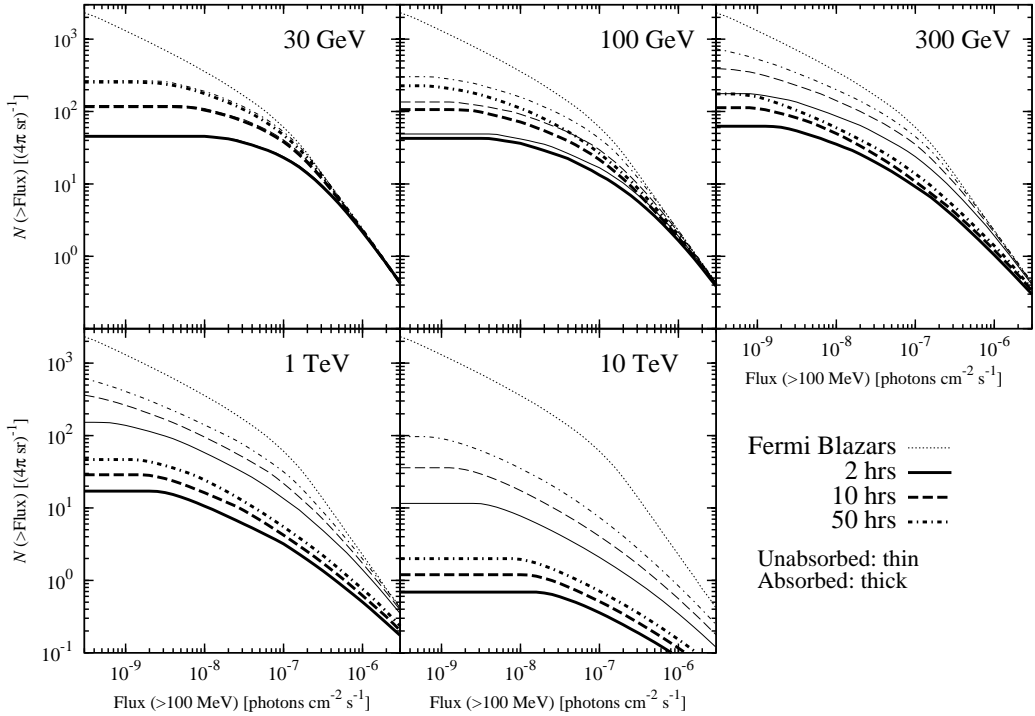} 
\end{center}
\caption{The source counts of {\it Fermi} blazars which are detectable
  by CTA, as a function of flux in the {\it Fermi} band ($>$100
  MeV). Five panels are for five photon energy bands of CTA, and three
different curves are for the 
  different CTA flux sensitivities for each blazar that are the same as those
in Fig. \ref{fig:z_dist}.
See Table \ref{tab:sens} for the sensitivity limits in
physical units. 
The dotted curves are for all Fermi blazars 
  regardless of the detectability by CTA. Thick curves take into account the intergalactic
  absorption, while thin curves do not.\label{fig:f_count_fermi} }
\end{figure*}

Although $\sim$1000 blazars will be detected by {\it Fermi} in all
sky, a simple systematic follow-up of all these blazars will not be
practical for CTA. In a realistic future observation, we should set
appropriate threshold flux and redshift range of {\it Fermi} blazars
to efficiently select the follow-up targets for CTA.
Fig. \ref{fig:f_count_fermi} shows the number of blazars detectable by
CTA as a function of Fermi GeV flux, for three different sensitivities
of CTA observation.  This figure tells us that the fraction of {\it
  Fermi} blazars detectable by CTA becomes smaller for higher photon
energy bands due to sensitivity and intergalactic absorption,
indicating that a follow-up survey may become inefficient if only the
flux information is utilized.

\begin{figure*}[t]
\begin{center}
\FigureFile(140mm,140mm){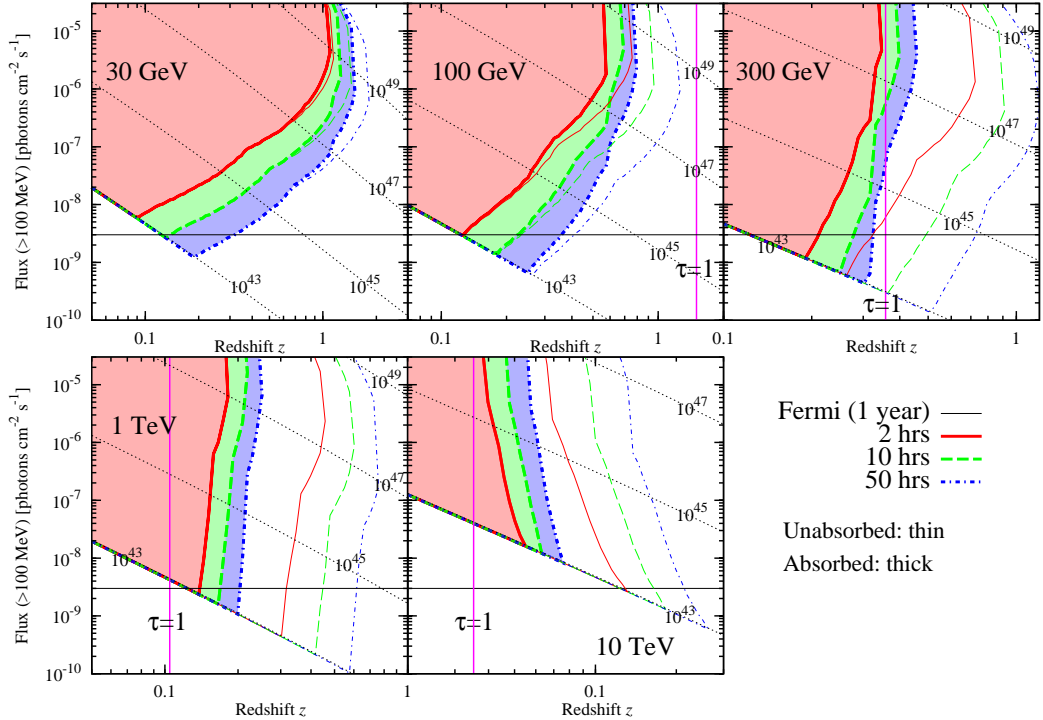} 
\end{center}
\caption{ The region of {\it Fermi} blazars that can be detected by
  CTA, in Fermi photon flux and redshift plane.  The five panels
  correspond to different photon energy bands of CTA, as indicated in the
  panels.  Different colors correspond to different CTA flux
  thresholds that are the same as those in Figs. \ref{fig:z_dist} and
  \ref{fig:f_count_fermi}.  The regions detectable are shaded by the
  corresponding colors.  The horizontal black solid line shows 1-year
  Fermi sensitivity, and the vertical magenta solid line
  is the redshift at which the optical depth of intergalactic
  absorption becomes unity, i.e. $\tau(z_\tau) = 1$.  
  Dotted black lines are contours of
  gamma-ray luminosities ($\nu L_\nu$ at restframe 100 MeV) of blazars 
  in units
  of [erg/s] as indicated in panels.
\label{fig:survey}
}
\end{figure*}

Since $\sim$70 \% of {\it Fermi} blazars have already been measured their 
redshifts ({Abdo} {et~al.} 2010), it is expected that a significant fraction of Fermi blazars will have
their redshift information. Therefore,
the redshift information may also be useful to select Fermi blazars as
the targets for CTA observations, though using redshift information
might introduce a further bias in the resulting sample.
Fig. \ref{fig:survey} shows the region in the Fermi flux versus
redshift plane for blazars than can be detected by CTA for three
different sensitivities. We denote $z_\tau$ (depending on observed gamma-ray energy
$E_\gamma$) as the redshift at which the absorption optical depth
becomes $\tau(z_\tau, E_\gamma)=1$.  It should be noted that, even for the same
Fermi flux, higher-$z$ blazars are more difficult to detect in VHE
bands. This is not only by the effect of intergalactic absorption;
another effect is that VHE flux becomes relatively smaller compared
with the Fermi band at larger distances, because of the larger
absolute luminosity and the assumed SED sequence.
If the SED sequence is valid, one must be careful to discriminate
between these two effects in the future analyses. 

These two figures will be useful to design the follow-up strategy of
{\it Fermi} blazars by CTA (e.g., the {\it Fermi} flux and redshift
thresholds for CTA targets). The spectral index in the Fermi band may
also be useful for efficient target selection.  A variety of target
selection strategies are possible, and the best observing strategy
must be determined according to the scientific purposes.

\section{On the determination of the EBL}
\label{sec:dis}

Now we consider to measure EBL by future VHE observations of blazars
by CTA, as a particular application of our result.  An obvious
approach is to use the brightest blazars and measure their VHE spectra
precisely, to measure the intergalactic absorption feature and hence
EBL.  The improved CTA sensitivity would allows us to measure the
absorption features in a wider range of photon energy, and
correspondingly, redshift.  The optical depth to a source at a
redshift $z$ is an integration of EBL from redshift zero to $z$ along
the photon path, and the optical depth is mainly contributed by EBL
photons whose frequency satisfies the relation $E_\gamma h\nu_{\rm
  EBL} \sim 2 (m_e c^2)^2$.  Therefore EBL measurements by blazars
having a variety of redshifts would give us information about
evolution of EBL as well as spectrum.

Another possible approach to measure EBL is using a break in redshift
distribution by intergalactic absorption.  Although this approach
would require a larger sample than using a single VHE spectrum of
bright blazars, the uncertainty concerning the intrinsic spectrum
could be minimized by looking a statistical signature in redshift
distribution.  Here we quantitatively discuss the feasibilities of
these two approaches, which are expected to be complementary to each
other.

\subsection{Absorption Features in Brightest Blazar Spectra}

\begin{figure}
\begin{center}
\FigureFile(90mm,90mm){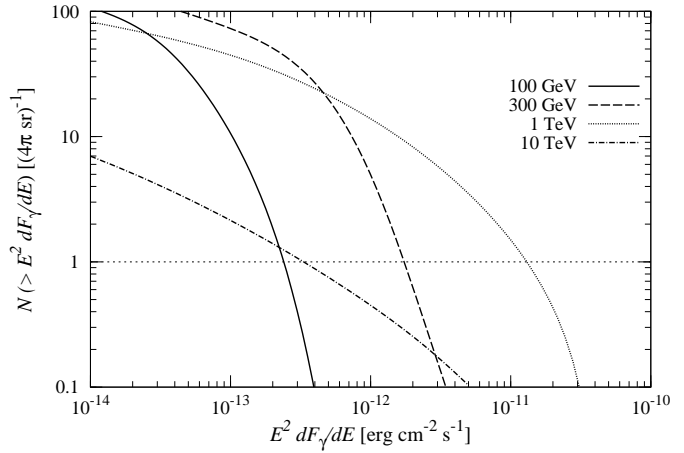} 
\end{center}
\caption{Source counts of VHE blazars as a function of
VHE flux in the four energy bands indicated in the figure,
assuming a follow-up of
{\it Fermi} blazars down to the flux threshold of
 $F(>{\rm
  100 MeV}) =3\times10^{-9} \ \rm photons \ cm^{-2} \ s^{-1}$.
Here, the redshift range is limited 
around the expected break by
EBL absorption: $z_\tau/2<z<2\times z_\tau$. 
The intergalactic absorption is taken into account. 
Horizontal dotted line marks the level of one blazar
in the entire sky.
\label{fig:f_count_ztau} 
}
\end{figure}

Fig. \ref{fig:f_count_ztau} shows blazar source counts as a function
of VHE flux around $z_\tau$ in the entire sky for the four energies of
100 GeV, 300 GeV, 1 TeV, and 10 TeV, in the case of following up {\it
  Fermi} blazars with the intergalactic absorption taken into account.
The values of $z_\tau$ are 1.5, 0.36, 0.10, and 0.035 for each energy,
respectively.  The flux of the brightest blazars available at the four
photon energy bands are $\sim$ 2, 20, 100, and 3 $\times 10^{-13} \rm
erg \, cm^{-2} s^{-1}$ (the fluxes at which the expected number
becomes order unity in this figure).

First we set the required signal-to-noise to be 5$\sigma$ per
logarithmic energy bin width of $\Delta E/E = 1$ to detect absorption
feature.  
This corresponds to 7.2 $\sigma$ detection for integrated flux.  From the
estimates of brightest blazar flux, the required observing time to
achieve this $S/N$ is 450, 0.36, 0.003, and 40 hours for the four
energy bands, respectively.  The photon energy resolution of CTA may
become as good as $\Delta E/E$ to be 0.1, and if we require $S/N=5$
per this spectral resolution, the required significance for integrated
flux becomes 19 $\sigma$. Then
the required observing time becomes 3000, 2.4, 0.02, and 270
hours for the four energy bands, respectively. 
These results indicate that we will
have bright blazars at various redshift range of 0.01--1.5 to measure
EBL with a reasonable amount of observing time. Especially, we will be
able to obtain high resolution spectra of bright blazars with
reasonable observing time between 0.3--1 TeV (corresponding to $z=$0.10--0.36).

\subsection{Cut-offs in Redshift Distributions}
\begin{figure}[t]
\begin{center}
\FigureFile(90mm,90mm){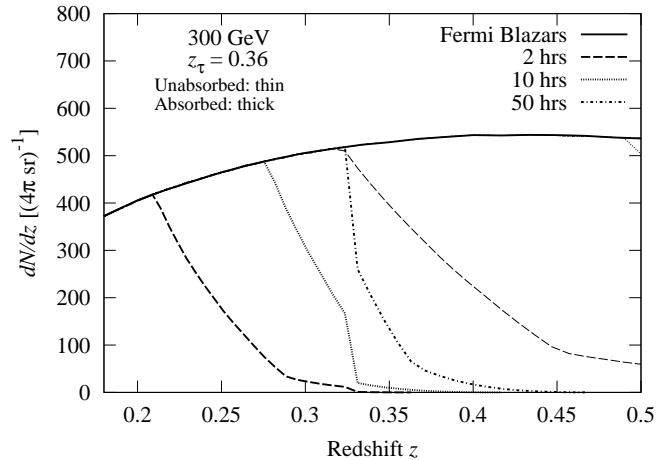} 
\end{center}
\caption{ 
The redshift distribution of blazars
detectable in the 300 GeV band, 
for the case of following
  up {\it Fermi} blazars down to
the Fermi flux threshold of $F(>{\rm
  100 MeV}) =3\times10^{-9} \ \rm photons \ cm^{-2}\ s^{-1}$. 
Different three curves are for different CTA sensitivity limits
as indicated in the figure in terms of the exposure time.
(See Table \ref{tab:sens} for the sensitivity flux in physical units.)
  Thin and thick curves correspond to unabsorbed cases and absorbed cases,
  respectively. Redshift distribution of {\it Fermi} blazars is also
  shown as black solid line. \label{fig:z_dist_tau} 
}
\end{figure}

Next we consider the break signature in redshift distribution.  Here
we suppose that redshifts of the majority of Fermi blazars are already
known at the time of future VHE observation, which seems a reasonable
assumption as discussed in the previous section.  Suppose a VHE photon
energy $E_\gamma$ and corresponding $z_\tau$.  We expect a strong
break around $z_\tau$ in the redshift distribution of Fermi blazars
that are detectable by the VHE photon energy band around $E_\gamma$,
and such a break should give a strong constraint on EBL.  Such a break
is demonstrated in Fig. \ref{fig:z_dist_tau}, where we show the
redshift distribution of Fermi blazars whose Fermi flux is brighter
than $F(>{\rm 100 MeV}) =3\times10^{-9}\ \rm photons \ cm^{-2}\
s^{-1}$ and which are detectable at 300 GeV band with several
different CTA sensitivities. Most of Fermi blazars at $z \lesssim 0.3$
can be detected by CTA, while sharp cut-offs appear as expected, at
redshifts significantly lower than the cases ignoring intergalactic
absorption.

\begin{figure*}[t]
\begin{center}
\FigureFile(150mm,150mm){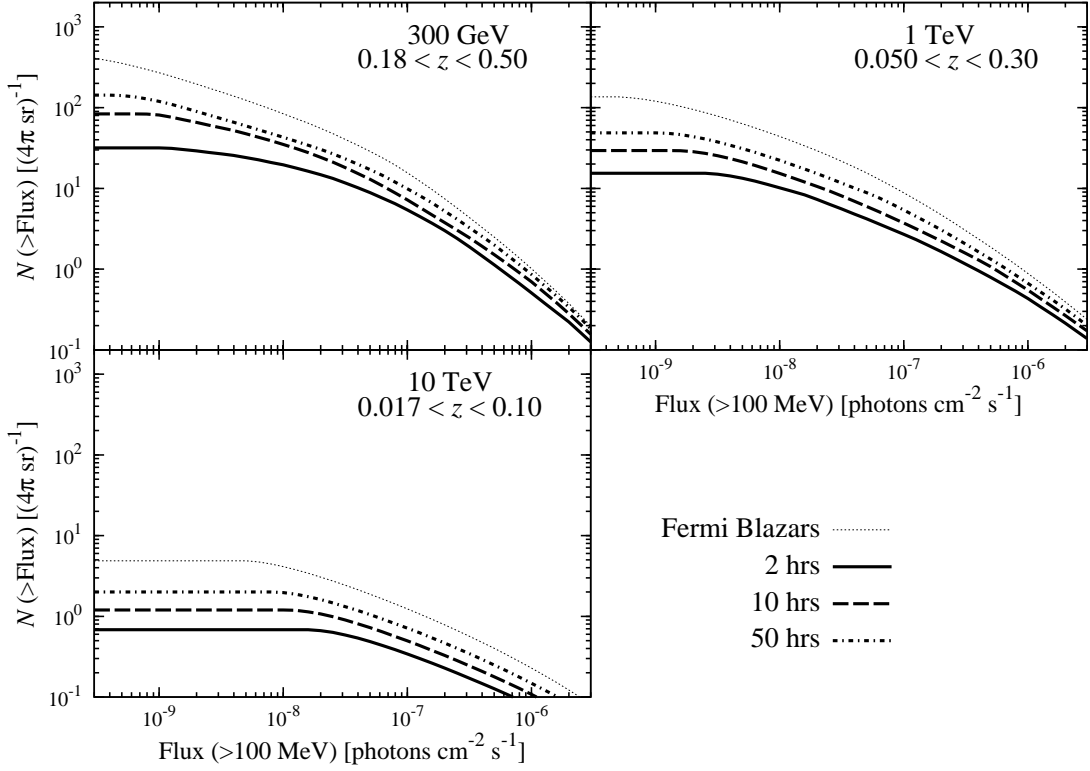} 
\end{center}
\caption{
Source counts of Fermi blazars as a function of Fermi-band flux,
within limited redshift ranges around the expected redshift
break by EBL absorption. Different panels are for different
VHE photon energy bands, and three different curves in each panel
are for those detectable by the VHE band with the CTA exposure
time indicated in the figure. See Table \ref{tab:sens} for 
the corresponding VHE flux sensitivity in physical units.
The dotted line is the original Fermi source counts regardless
of the detectability in VHE bands. \label{fig:f_count_ebl}
}
\end{figure*}

The key question for this approach is whether there are a sufficient
number of blazars to construct a statistically large enough sample to
see the break.  This available number should change with the supposed
redshift (or VHE photon energy).  Fig. \ref{fig:f_count_ebl} gives an
answer to this question; it shows blazar counts as a function of Fermi
flux, for Fermi blazars around $z \sim z_\tau$ in three VHE photon
energy bands.  In this figure, we show source counts of blazars that
can be detected with several different CTA sensitivities, as well as
the original Fermi source counts. Then we can estimate the number of
targets to be observed and detection rate at VHE bands for a given
Fermi threshold flux and VHE sensitivities.  We did not show panels
for 30 and 100 GeV because the expected break redshift by
intergalactic absorption is comparable with or larger than the
redshift limit coming from CTA sensitivity with a reasonable amount of
observing time (see Fig. \ref{fig:survey}).

>From this result, it seems difficult to construct a statistically
large ($\gtrsim$ 10) sample at 10 TeV. However, in the 300 GeV and 1
TeV bands, we will be able to construct a sample of a few tens of
blazars with 50 and 30 hrs of total observational time by following up
Fermi blazars down to $F(>{\rm 100 \, MeV}) = 3\times10^{-9}\ \rm
photons \ cm^{-2}\ s^{-1}$, respectively. Here we assumed that all
samples are 5 $\sigma$ detection in integral flux and the minimum
observational time for each blazar is 0.5 hrs. If we require
10$\sigma$ detection for each blazar at these VHE sensitivities, the
total observation time necessary for this survey becomes 200 and 100
hrs, respectively. Therefore, this statistical approach seems feasible
in the VHE energy range of 0.3--1 TeV, which is complementary to EBL
measurements using spectral break of bright blazars.

It should be noted that the location and shape of the break in the
redshift distribution depends not only on EBL but also VHE luminosity
function of blazars, which may induce some systematic uncertainties in
the EBL measurement by this approach.  However, we should be able to
construct luminosity distribution of VHE blazars that is not affected
by intergalactic absorption by using blazars slightly below $z_\tau$.
We do not expect a strong cosmological evolution of VHE luminosity
function of blazars in a small range of redshift around $z_\tau$, and
we can apply the VHE luminosity distribution below $z_\tau$ to derive
the optical depth of intergalactic absorption,
without invoking uncertain theoretical modeling.

\section{Summary}
\label{sec:sum}

In this paper, we estimated the expected source counts and redshift
distribution of VHE blazars for the next generation IACTs such as CTA
and AGIS missions based on the latest blazar GLF model of
{Inoue} \& {Totani} (2009). For this purpose, we developed a new SED
sequence formula of blazars taking into account the latest VHE data,
while the previous sequence formulae were constructed using only data
at photon energies below the EGRET/Fermi energy band. The parameters
of our blazar GLF are also refined with this new SED formula, by
fitting to the GeV blazar data.  Our modeling does not include time
variabilities of VHE blazars, and blazars at flaring states would be
more easily detected than estimated here.

We made predictions for future VHE blazar survey in two observing
modes: one is a blind survey in a blank field, and another is a
following up survey of {\it Fermi} blazars. We found that CTA will
detect a few VHE blazars by a blind survey using a total survey time
of 100 hours.  Therefore a large amount of observing time ($\gtrsim
1000$ hrs) is required to construct a statistically large sample of
blazars selected only by VHE bands.  However, this suggests that
blazar contamination in the Galactic plane survey should not be
significant even in the era of the next generation IACTs.  We also
found that future all sky gamma-ray detectors such as HAWC and
Tibet-III/MD, will detect only a few VHE blazars in one year survey in
the entire sky.  The survey design for a follow-up survey of Fermi
blazars should be dependent on the scientific purposes. Here we
presented a plot for regions in the Fermi flux versus redshift plane
where the Fermi blazars can be detected by VHE observations for
several different sensitivities.  

As a particular example of Fermi blazar follow-up surveys, we
considered a survey for the purpose of determination of EBL by VHE
observation. CTA can observe VHE blazars that are sufficiently bright
to get detailed spectra with high S/N in the redshift range of $z
\sim$ 0.10--0.36, corresponding to the absorption cut-off energy
1--0.3 TeV, and hence we can constrain not only EBL flux but also its
spectra and/or redshift evolution.  It will also be possible to
construct a statistically large sample ($\gtrsim 30$) of blazars at $z
\sim $ 0.10--0.36 to constrain EBL by the sharp break in the redshift
distribution. This approach could avoid or minimize the uncertainty
about intrinsic blazar spectra, and hence could be complementary to
using a few spectra of brightest blazars.

This research has made use of the NASA/IPAC Extragalactic Database
(NED) which is operated by the Jet Propulsion Laboratory, California
Institute of Technology, under contract with the National Aeronautics
and Space Administration. This work was supported by the Grant-in-Aid
for the Global COE Program "The Next Generation of Physics, Spun from
Universality and Emergence" from the Ministry of Education, Culture,
Sports, Science and Technology (MEXT) of Japan. YI acknowledges
support by the Research Fellowship of the Japan Society for the
Promotion of Science (JSPS).

\appendix
\section{The New Blazar SED Sequence Templates}

We introduce some modifications on the inverse Compton (IC) component
of the blazar SED sequence model of IT09 to make it in better
agreement with VHE blazars data. We define $\psi(x) \equiv \log_{10}
[\nu L_\nu / (\rm erg \ s^{-1})]$ with $x \equiv \log_{10} (\nu / \rm
Hz)$ ($\nu$ in rest-frame). The empirical SED sequence model of
blazars is the sum of the synchrotron $[\psi_s(x)]$ and IC
$[\psi_c(x)]$ emissions. Each component is described by a combination
of a linear and a parabolic function at low and high photon
frequencies, respectively.  We take $\psi_R \equiv \log_{10} [L_R /
(\rm erg \ s^{-1})]$ as a reference of a blazar luminosity, where
$L_R$ is $\nu L_\nu$ luminosity in the radio band ($\nu_R = $5 GHz or
$x_R = 9.698$).

Here we only describe the modified
points from the IT09 blazar sequence model. The peak frequency of
IC component $\nu_c$ is determined by the relation to that of the
synchrotron component, $\nu_s$,
where $\nu_s$ has been determined as a function of
$\psi_R$ as in IT09.  This relation has been changed into the
following equation:
\begin{equation}
\nu_c/\nu_s=\left\{ \begin{array}{ll}
    	 5\times 10^8 & (\psi_R < 43.0) \\
    	5\times 10^8 [10^{(\psi_R-43.0})]^{-0.1} & (\psi_R \ge 43.0) \\
    \end{array}
    \right. , 
\end{equation}
instead of the fixed value $\nu_c/\nu_s = 5 \times 10^8$ used by IT09. 
The parabolic part of the IC component, $\psi_{c2}$ is modified
as follows:
\begin{equation}
\psi_{c2}(x)\equiv\left\{ \begin{array}{ll}
    	-[(x-x_c)/\sigma]^2+\psi_{c,p} & (x<x_c) \\
    	-1.5[(x-x_c)/\sigma]^2+\psi_{c,p} & (x \ge x_c) \\
    \end{array}
    \right. ,
\end{equation} 
from eq. A6 of IT09. Note that the shape of the parabolic part is now
different for the synchrotron and IC components.  Such a difference is
possible by several effects, e.g., external photon field for target
photons of IC, internal absorption of high-energy gamma-rays by pair
production, or the Klein-Nishina effect.  Finally, the peak luminosity
of the IC component, $\psi_{c,p}$, is changed into the following form:
\begin{equation}
\psi_{c,p}=-0.014(\psi_R-36.2)(\psi_R-44.6)(\psi_R-55.0)+47.7.
\end{equation}
from eq. A13 of IT09. 

\begin{figure}[t]
\begin{center}
\FigureFile(80mm,80mm){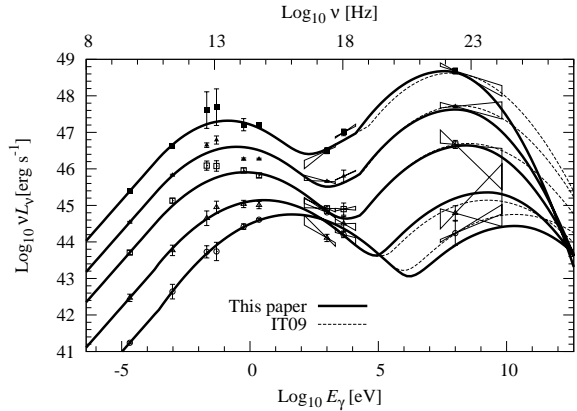} 
\end{center}
\caption{The blazar SED sequence. The data points are the average SED
  of the blazars studied by {Fossati} {et~al.}  (1998) and {Donato} {et~al.} (2001). The
  solid and dashed curves are the empirical SED sequence models of this paper and IT09, respectively.\label{fig:sed} }
\end{figure}

Fig. \ref{fig:sed} shows the blazar sequence SED of IT09 and of this paper.

\bibliography{}
\bigskip

\end{document}